# How Periodic Forecast Updates Influence MRP Planning Parameters: A Simulation Study

Wolfgang Seiringer[a], Klaus Altendorfer[a], Thomas Felberbauer[b], Balwin Bokor[a], Fabian Brockmann[c]

[a]*Department for Production and Operations Management,*
*University of Applied Sciences Upper Austria, A-4400 Steyr, Austria,*
*Wolfgang.Seiringer@fh-steyr.at, Klaus.Altendorfer@fh-steyr.a, Balwin.Bokor@fh-steyr.at*
[b]*Department of Media and Digital Technologies,*
*University of Applied Sciences St. Pölten, A-3100 St. Pölten, Austria*
*Thomas.Felberbauer@fh-stp.ac.at*
[c] *Department of Business and Management Science,*
*NHH Norwegian School of Economics, N-5045 Bergen, Norway,*
*Fabian.Brockmann@nhh.no*

**Abstract**

In many supply chains, the current efforts at digitalization have led to improved information exchanges between manufacturers and their customers. Specifically, demand forecasts are often provided by the customers and regularly updated as the related customer information improves. In this paper, we investigate the influence of forecast updates on the production planning method of Material Requirements Planning (MRP). A simulation study was carried out to assess how updates in information affect the setting of planning parameters in a rolling horizon MRP planned production system. An intuitive result is that information updates lead to disturbances in the production orders for the MRP standard, and, therefore, an extension for MRP to mitigate these effects is developed. A large numerical simulation experiment shows that the MRP safety stock exploitation heuristic, that has been developed, leads to significantly improved results as far as inventory and backorder costs are concerned. An interesting result is that the fixed-order-quantity lotsizing policy performs—in most instances—better than the fixed-order-period lotsizing policy, when periodic forecast updates occur. In addition, the simulation study shows that underestimating demand is marginally more costly than overestimating it, based on the comparative analysis of all instances. Furthermore, the results indicate that the MRP safety stock exploitation heuristic can mitigate the negative effects of biased forecasts.

## 1. Introduction

The increasing digitalization of production systems and supply chains improves the information exchange between supply chain partners, i.e., between the supplier (manufacturer) and the customer. As supply chain literature has shown, timely demand information and its updates improve the supply chain performance. Therefore, a typical production system setting is that customers provide forecasts for their demands with a long planning horizon and regularly update these forecasts. An appropriate method to model such behaviour is the forecast evolution model known from Heath and Jackson (1994), which has been applied in Norouzi and Uzsoy (2014) and Altendorfer and Felberbauer (2023) to evaluate forecast accuracy. From a practical perspective, MRP (Material Requirements Planning) is still an important production planning method (Seiringer et al. 2022a) and (Louly and Dolgui 2013); however, it assumes a deterministic demand setting where set customer demand remain unchanged. Many studies show that the performance of MRP depends very much on the planning parameter setting used as in Molinder (1997), Jodlbauer and Huber (2008) and Altendorfer (2019). These planning parameters are production lotsize, planned lead time and safety stock for each item (i.e. material). The MRP algorithm is based on a deterministic production system and demand assumptions; however, to react on the stochastic nature of a production system and expected demand updates, MRP is calculated on a regular basis with overlapping planning horizons, i.e., a rolling horizon planning is applied. On the one hand, this rolling horizon planning enables the incorporation of new and updated demand information and production system feedback (e.g., production orders being finished before or after the planned due date). On the other hand, this approach is one main driver for MRP system nervousness as stated by Ho and Ireland (1998) and Li and Disney (2017). Since forecasts are regularly updated and MRP planning regularly calculates new production plans, i.e., production orders, an interaction between the forecast quality and the MRP planning behaviour is observable. In this study, MRP performance is evaluated in terms of cost-related impacts, considering inventory holding and backorder costs to ensure a consistent assessment of planning parameter settings and forecast dynamics. Furthermore, the service level is applied for certain in-depth discussions regarding the behaviour of the planning methods. As the MRP performance depends on the respective planning parameters, we study the interrelation between optimized planning parameters, i.e., by conducting a simulation study with solution space enumeration, and different forecast evolution behaviours, i.e., demand quantity uncertainty is addressed for a multi-item, multi-stage production system, with deterministic processing times and stochastic setup times. Specifically, for the lotsizing decision, multiple approaches are available. However, we limit our study to Fixed-Order-Quantity (FOQ) and Fixed-Order-Period (FOP) lotsizing policies, since they are regularly evaluated in literature (see Enns (2002) and Gansterer et al. (2014) for FOQ and Altendorfer and Felberbauer (2023) for FOP) as and are widely applied in practice. Research on MRP, such as Altendorfer and Felberbauer (2023) or Bregni et al. (2013), highlights that forecast updates can cause significant disturbances in production systems. Our findings confirm this, showing that rolling horizon planning with forecast updates frequently triggers short-term production orders. In particular, when using the FOP lot-sizing policy, these short-term orders are often generated with low lotsizes, increasing system nervousness.

Therefore, a simple short-term safety stock exploitation heuristic is developed which extends the MRP standard calculation and allows safety stocks to be applied for short-term demand uncertainties. Note that safety stocks are often used by production planners for manual intervention to react on short-term demand quantity updates. However, the extension that has been developed is integrated in the MRP algorithm, i.e., no manual effort is required. Specifically, the following research questions are addressed in this paper:

- RQ1: How does uncertainty in demand forecast updates impact the optimization of MRP parameters when using a Fixed-Order-Quantity (FOQ) or Fixed-Order-Period (FOP) lot-sizing policy?
- RQ2: What is the increase in MRP performance in a rolling horizon MRP planned production system with forecast updates if the developed safety stock exploitation heuristic is applied?
- RQ3: What is the influence of temporary and permanent forecast bias on the optimized MRP planning parameters and the respective costs when forecast updates occur?

Research question RQ1 addresses an MRP standard setting wherein long-term forecasts are provided by the customers and respective demand quantity updates occur, when we assume that all of them include an information gain. Note that in this paper, the forecast updates are always related to demand quantity updates, and no other effects—such as demand shifting or changes of demand due dates—are addressed. The respective results contribute to a better understanding of MRP behaviour and provide some managerial implications on how to parameterize this planning method in its standard form to mitigate the negative effects of updates. The extension developed and discussed in RQ2 provides a simple heuristic to improve MRP with respect to forecast updates, and then the respective cost improvements are evaluated. From a managerial perspective, this simple extension can have a high impact on the performance of the production system. Furthermore, its generic formulation and simple integration into the MRP standard algorithm foster its practical applicability. In research question RQ3, different cases where forecasts are biased, or not all forecast updates include an information gain (i.e., temporary forecast bias), are analysed, which contribute to a better understanding of how information rationing and distortion influence the MRP performance and the respective optimized planning parameters.

The rest of this article is structured in the following manner: initially, a review of literature related to forecast evolution is presented. This is followed by a section detailing the production system under study, along with the forecasting models applied and the MRP safety stock exploitation heuristic developed. Subsequently, the setup and findings of the numerical study are presented and analysed. The article concludes with a summary of the findings and an outlook on future research possibilities.

## 2. Related Literature

This section presents a literature review on the impact of forecast uncertainty on production systems. It explores the general effects of forecast uncertainty and examines existing strategies for managing these uncertainties through adjustments to production parameters. Specifically, the focus is on MRP systems, with particular attention to the key parameters of planned lead time, lotsize, and safety stock.

### *2.1 Forecast Uncertainty in Production Planning*

Forecast uncertainty in production planning systems has been extensively researched, with Whybark and Williams (1976) identifying two types of forecast uncertainty: forecast quantity uncertainty and demand timing uncertainty. Recent literature has distinguished between unsystematic and systematic forecast behaviours, wherein Zeiml et al. (2019) and Altendorfer and Felberbauer (2023) further differentiated between systematic behaviours into overbooking and underbooking scenarios. Altug and Muharremoglu (2011) demonstrated the significance of high-quality demand information and long planning horizons for optimal performance of production planning systems. Altendorfer et al. (2016) suggested that utilization levels significantly below 100 % are necessary to deal with demand uncertainty and that overall costs increase with higher demand uncertainty. Wijngaard (2004) demonstrated that demand information effectively reduces safety stock and the bullwhip effect, yet its significance is constrained by factors such as capacity tightness, forecast horizon, and the uncertainty inherent in demand information. In terms of short-term versus long-term demand information, Iida and Zipkin (2006) showed that short-term demand information is more relevant for production planning, but a longer planning horizon is beneficial to deal with trends and cycles in long-term forecasts. Herding and Mönch (2023) enhanced production planning by integrating short-term demand information into a mixed-integer linear programming model, termed the STDSM approach. This method, combined with master and allocation planning within a rolling horizon framework, was validated through simulation in a simplified semiconductor supply chain. Enns (2002) demonstrated the need for production planning systems to adapt to biased forecast information to achieve a certain level of service. While van Kampen et al. (2010) did not focus on biased forecast situations, they investigated the varying effects of different types of demand uncertainty on production planning. Within this context, Rötheli (2018) underscore the strategic advantage of business cycle forecasting, showing how it aids firms in optimizing output mix in the short run and adjusting operational levels in response to market cycles. Overall, these studies emphasize the importance of accurate demand information and planning horizons to mitigate the effects of forecast uncertainty on production planning systems.

Despite increased inventory levels and backorder costs resulting from forecast uncertainty, researchers found out how information sharing can reduce these effects. Wu and Edwin Cheng (2008) showed that by sharing information in a multi-echelon supply chain with uncertain demand, safety stock levels can be reduced and performance enhanced. Huang et al. (2003) found that sharing information in a two-stage supply chain can reduce lead times and inventory levels, and also lead to improved performance. Lee and Whang (2000) also showed how shared information can improve the performance of multiple partners in a supply chain. They also discussed different models for information sharing. Overall, these studies highlight the benefits of information sharing in supply chains, enabling a more efficient flow of materials and information, optimizing production processes, and enhancing performance by providing visibility and transparency across the supply chain.

*2.2 MRP Parameter Setting*

Since MRP is a frequently used planning method for production systems, a lot of literature focuses on the optimization of the MRP parameter levels. This includes safety stock, planed lead time, and lotsize.

**2.2.1. Safety Stock**

Starting with adapting only the safety stock, Seiringer et al. (2022b) demonstrated that for unsystematic forecast behaviours, an increase in uncertainty leads to an increase in safety stock levels, resulting in reduced overall costs. Molinder (1997) established a positive correlation between higher variabilities in the system and the higher safety stock levels required to maintain a certain level of service. Enns (2002) showed that the marginal benefit of adding more safety stock diminishes as the level of safety stock and service levels increase. Schoenmeyr and Graves (2009) applied safety stocks to decouple individual nodes within the supply chain. In addition, Boulaksil (2016) demonstrated an increasing tendency to place safety stock downstream in the supply chain with rising demand uncertainty. Furthermore, Manary and Willems (2008) developed an adapted technique to determine the appropriate safety stock levels in biased forecast situations by adjusting the desired service level parameters. Whereas, the utility of safety stock at the master production schedule (MPS) level as a method to mitigate schedule nervousness (hedging the MPS) in an MRP planned production system was explored by Sridharan and Lawrence LaForge (1989). They concluded based on experimental analysis, that while modest amounts of safety stock can enhance schedule stability and reduce cost errors, excessive safety stock may exacerbate schedule nervousness, suggesting the need for alternative strategies to address forecast inaccuracies.

**2.2.2. Planned Lead Times**

Another series of papers examines how to set planned lead times under forecast uncertainty. According to Enns (2001), a potential solution to mitigate the impact of demand uncertainty is to increase the planned lead time. However, the most effective approach may vary, depending on the type of uncertainty. Also, Altendorfer (2019) found higher planned lead times to be advantageous for mitigating forecast uncertainty when shop loads are high. Enns (2002) showed that in scenarios with biased and unbiased forecasts, a longer planned lead time can lead to a higher service level. It is important to note that the lack of forecast evolution in the results of Enns (2002) has an effect similar to that of Enns (2001), where inventory levels increase to mitigate uncertainty.

**2.2.3 Interrelation between Safety Stock and Planned Lead Times**

While one can set a planned lead time or a safety stock to face forecast uncertainty, it is also possible to exchange them. Whybark and Williams (1976) suggested that for demand quantity uncertainty, safety stock is beneficial, while for demand timing uncertainties, safety lead time is preferable in single-product, single-stage production environments. However, this approach may not be suitable for multi-product, multi-stage production systems, as discussed in Enns (2002). Dong and Lee (2003) suggested that higher planned lead times may be preferable in situations with higher demand timing uncertainty, but the results are inconclusive. In systematic biased demand uncertainty situations, Altendorfer (2019) and Buzacott and Shanthikumar (1994) indicated that a higher underbooking behaviour tends to result in less planned lead time and more safety stock. Again, van Kampen et al. (2010) showed that in negative biased forecast situations with low and high levels of demand quantity uncertainty, a safety lead time leads to a higher service level at a comparable inventory level, which is different from the results of Altendorfer (2019) and Buzacott and Shanthikumar (1994), who suggested increasing safety stock levels. Enns (2002) showed that in underbooking scenarios, setting safety stock is preferable to setting a safety lead time, which is in line with Altendorfer (2019) and Buzacott and Shanthikumar (1994). However, there is no clear evidence from Enns (2002) to determine if this also holds true for overbooking scenarios. In addition, Molinder (1997) found that in the case of low lead time variability, the use of safety stock instead of safety lead time leads to the lowest costs. Overall, these studies provide insights into the complex interrelation between demand uncertainty, safety lead time, and safety stock in production planning systems.

**2.2.4. Lotsize**

Apart from safety stock and lead time adjustment, it is also possible to adapt the lotsize to increase the system performance. Callarman and Hamrin (1979) and (1984) found that as demand uncertainty increases, the influence of setup and holding costs decreases, which is further supported by Wemmerlöv (1989). However, these studies did not integrate setup times or a simulation of shop floor behaviour. But Enns (2001)—who investigated the effects of lotsizes and shop floor behaviour on tardiness – found that both small and large lotsizes can result in high tardiness. Also, Enns (2002) showed that there is an optimal lotsize that minimizes tardiness. In addition,

Altendorfer (2015) suggested that large lotsizes lead to an additional safety stock, which can be further increased by safety stock and safety planned lead time to provide a higher service level.

**2.2.5. Joint Parameter Setting**

The literature on the simultaneous optimization of all three MRP parameters is limited. Vaughan (2006) provided an analysis of how the MRP parameters, namely safety stock, planned lead time, and lot sizing, are interdependent. Gansterer et al. (2014) offered six simulation-based optimization approaches for the three MRP parameters in different demand market situations, with Variable Neighborhood Search (VNS) emerging as the most effective for exploring the parameter space. They highlighted the strong interdependence of these parameters. Enns (2001) investigated the effect of setting planned lead time and lot sizing simultaneously on the mean tardiness and work-in-progress (WIP). Larger lotsizes generate an extra safety stock that reduces the requirement for additional safety stock. However, larger lotsizes also produce longer realized lead times and greater variability, which necessitates more safety stock or a longer planned lead time. This finding was also supported by Altendorfer (2019) and Seiringer et al. (2022a); however, a very simple single-stage production system was studied.

*2.3. Contribution to Literature*

Our literature review shows that the influence of demand uncertainty on production planning is an interesting and widely studied topic. Table 1 presents the reviewed literature, including the analysed MRP parameter settings and the type of demand uncertainty. The current paper addresses the MRP parameters safety stock, planned lead time and lotsize when a quantity uncertainty occurs. Looking at the reviewed literature shows, that the same settings are studied in Enns (2001), Molinder (1997), and (Seiringer et al. 2022a). However, none of these studies includes forecast updates which are the focus of this work.

Table 1: Overview of the reviewed literature.

| Author | Demand Uncertainty | | Planning Parameter | | |
|---|---|---|---|---|---|
| | Quantity | Timing | Safety Stock | Lead Time | Lotsize |
| Seiringer et al. (2022b) | x | | x | | |
| Schoenmeyr and Graves (2009) | x | | x | | |
| Manary and Willems (2008) | x | | x | | |
| Sridharan and Lawrence LaForge (1989) | x | | x | | x |
| Altendorfer (2019) | | x | x | x | x |
| Whybark and Williams (1976) | x | x | x | x | |
| Enns (2002) | x | | x | x | |
| Dong and Lee (2003) | | x | | x | |
| Buzacott and Shanthikumar (1994) | x | | x | x | |
| van Kampen et al. (2010) | x | | x | x | |
| Callarman and Hamrin (1979, 1984) | x | | x | | x |
| Wemmerlöv (1989) | x | | | | x |
| Altendorfer (2015) | x | x | | x | x |
| Vaughan (2006) | x | | x | x | x |
| Gansterer et al. (2014) | | x | x | x | x |
| Seiringer et al. (2022a) | x | x | x | x | x |
| Molinder (1997) | x | | x | x | x |
| Enns (2001) | x | | x | x | x |

Therefore, the specific effect of regular forecast updates remains an important and relevant research area, especially in light of the ongoing digitalization of production systems and supply chains. Currently, no study in the literature optimizes all three MRP planning parameters for a production system with demand updated according to a forecast evolution model.

In this context, our work makes two key contributions. First, we study the effect of forecast updates and their uncertainty on optimal production planning parameters, thereby contributing to the MRP parameter optimization literature. Second, we extend MRP by developing a simple yet practically applicable approach that enables production systems to benefit from uncertain information updates. This extension allows the system to mitigate the negative effects of disturbances related to these updates without manual intervention, providing both a theoretical and practical contribution to the MRP planning literature.

## 3. Production System and Forecast Modelling

In this section, the studied production system, the standard MRP planning method, the applied forecast evolution method, and the developed MRP safety stock exploitation heuristic are introduced. For all three elements – namely, the production system, planning method, and forecast modelling – a balance is struck between simplistic modelling and the incorporation of real-world effects and complexities.

### 3.1. Production System Structure and MRP Standard Planning

A multi-item, multi-stage production system serves as the basis for examining our research assumptions. The implementation was carried out in a discrete-event simulation model using AnyLogic. Since various sources of uncertainty are investigated and the interrelation of information updates with overlapping planning horizons is a key focus, purely analytical models are insufficient to capture all relevant aspects. Hence, simulation is applied. Simulation has been widely used to model the interrelation between forecast behaviour and production system performance, as demonstrated by Byrne and Heavey (2006), Enns (2002), (2011), Ali and Boylan (2011), Ali et al. (2012), Zeiml et al. (2019), and Seiringer et al. (2022b).

To facilitate a clear and structured analysis, we consider a production system that is sufficiently complex for in-depth analysis while allowing for clear result interpretability. It features a multi-machine and multi-stage structure, where multiple materials are processed on the same machine, and items require several machines for production. This setup balances real-world complexity with interpretability, ensuring that key interactions can be systematically analysed without excessive model complexity. Building upon this, we assume a system with eight final products that are produced on four different machines, along with two components manufactured on two additional machines. Figure 1 illustrates the bill of materials (BOM) and the corresponding routing information, detailing the structure of the production process. Drawing from the practical observation that products frequently need to be delivered on a regular basis, we operate under the assumption that all products should be delivered every fourth period. In detail, Products 10 and 14 have to be delivered in Periods {1, 5, 9, ….}, Products 11 and 15 in Periods {2, 6, 10, ….}, Products 12 and 16 in Periods {3, 7, 11, ….}, and Products 13 and 17 in Periods {4, 8, 12, ….}. Note that in the simulation model, the first demand occurrence is after 12 periods to enable the respective backward scheduling in MRP. As the backward scheduling and lotsizing (let us assume FOP lotsizing policy here for simplicity) lead to production orders being issued several periods in advance of their due date, including the demand of several periods, a number of forecast updates may occur between the issuance of the production orders of final products and their completion.

As shown in Figure 1, Products {10, 11, 12, 13} are first produced on Machine 102 and then on Machine 101, and use Component 20. Products {14, 15, 16, 17} are first produced on Machine 112 and then on Machine 111, and use Component 21. The expected order amount for all products is 800 pieces per demand period (day) and the processing time of the final products is 1.35 min/piece. Machines are available 24h/day which leads in combination with the periodic orders, the order amounts, and the processing times to a baseline shop load of 75% at Machines 101, 102, 111, and 112 without setup. Based on different setup times, three shop load situations are analysed. For products, the setup times {low=216 min/lot, medium=288 min/lot, high=331.2 min/lot} correspond to planned utilization levels of 90%, 95%, and 98% (for M101, M102, M111, and M112) when FOP 1 is applied to define the setup times used for the utilization scenarios. The expected value of setup time is equal for all products. These three utilization levels represent distinct scenarios that are systematically compared in a comprehensive study. Note the realized utilization in the simulation study depends on the applied lot-sizing policy and the resulting production orders. For components, the processing time is 0.68 min/piece, resulting in a baseline shop load of 75% without setup. The setup time is 94 min/lot, leading to a planned utilization of 88% (for M201 and M202) with FOQ 800 and 81.5% with FOQ 1600. The FOQ lot-sizing policies for components are selected based on preliminary simulation results to ensure lower utilization of component machines.

For generating production orders, the MRP standard is applied in our setting. This means that the four calculation steps: netting, lot-sizing, backward scheduling, and BOM explosion are applied, as described in Hopp and Spearman (2011). Production orders for both products and components are released exactly at the planned start date and no manual production planner intervention is modelled. In the case of delayed orders, the end date is calculated as the current date plus the planned lead time. In the present paper, the emphasis is on the impact of forecast updates. All processing times are modelled deterministically, and all setup times are lognormally distributed with a coefficient of variation of 0.2.

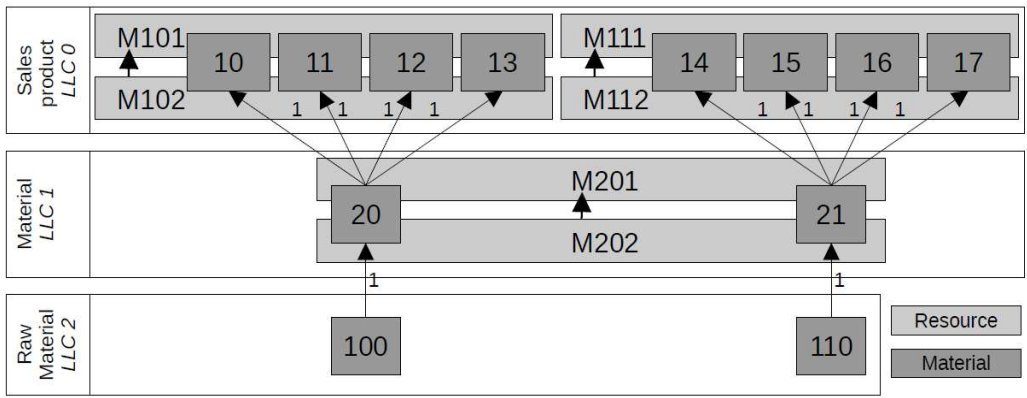

**Figure 1: Production system structure.**

### 3.2. MRP Safety Stock Exploitation Heuristic

A detailed analysis of the simulation results when the MRP standard is applied shows that demand forecast updates lead to an increased number of production orders, with small order amounts and immediate planned start dates. This implies a planning nervousness created through the combination of the netting step in MRP and the quantity updates as shown by Li and Disney (2017). Hence, a straightforward MRP safety stock exploitation heuristic to enable the use of available forecast information but avoiding the related nervousness is developed in this paper. To preclude this behaviour, the netting step of the MRP run is extended in the following manner. For final products, we track the due dates for which a production order has already been released to the production system, e.g., for a planned lead time of 2 periods and a FOP 3 policy (i.e., the demands of 3 periods are cumulated to one production lot), we release a production order in Period 1 for the net requirements of Periods 3–5. Then we track that for Periods 3–5 that a production order has already been issued. To avoid unnecessary additional production orders for these Periods 3-5, we modify the netting step in such a way that a new production order is not created if the projected-on-hand stock for the tracked periods falls below the safety stock (as in the MRP standard run; see Equation (1)), but only if the stock falls below zero (see Equation (2)). Note that for all periods for which no production order has yet been released, the netting step works as in an MRP standard, and the order is issued if the stock projected-on-hand falls below the safety stock. This leads to a setting where the safety stock can be applied for mitigating the negative effects of short-term forecast updates; however, the value of forecast information is still exploited in the MRP as short-term orders are still issued if the quantity update cannot be fulfilled from the safety stock. We call this the *MRP safety stock exploitation heuristic* as the available safety stock is exploited better in this setting.

For the MRP standard, the following netting calculation is performed for $n_t$ (net requirements in period $t$) with $s$ being the safety stock, $y_{t-1}$ being the stock in period $t-1$, $g_t$ being the gross requirements in period $t$, and $r_t$ being the scheduled receipts in period $t$.

$$n_t = \max\left(s - (y_{t-1} - g_t + r_t), 0\right) \text{ for } t > 0 \qquad (1)$$

When applying the safety stock exploitation heuristic, the following netting calculation applies, with $\delta$ being the period until which net requirements have already been aggregated to a production order, i.e., $\delta = 5$ in the example above.

$$\begin{aligned} n_t &= \max\left(-y_{t-1} + g_t - r_t, 0\right) \text{ for } t \leq \delta \\ n_t &= \max\left(s - (y_{t-1} - g_t + r_t), 0\right) \text{ for } t > \delta \end{aligned} \qquad (2)$$

Note that the current simulation time is assumed to be zero without loss of generality. A limitation of this heuristic is that less safety stock is available to react on reduced yield, e.g., if scrap or rework occurs in the production system. Even though this combination with scrap and rework could also be investigated, we leave this for further research, and focus only on the information update aspects in this paper.

### *3.3. Forecast evolution model*

In our study, customers provide their demand forecasts in accordance with the additive MMFE model (martingale model of forecast evolution) as shown in Heath and Jackson (1994), Güllü (1996) and Norouzi and Uzsoy (2014). This means that demand forecasts for final products are available for a long forecast horizon into the future and their quantity is periodically updated until the respective due dates. Such long-term forecasts can, for example, be based on agreement contracts Shen et al. (2019), which also imply regular orders, as introduced in Section 3.1. Note that the original MMFE model from Heath and Jackson (1994) also allows for the analysis of final product demand autocorrelation effects, which are not included in the current paper, i.e., we assume that the demand for each due date is independent of all other demands—the respective final product has no autocorrelation between due dates—and is independent of all other final products' demands (there is no autocorrelation between final products). However, Heath and Jackson (1994), Güllü (1996), and Norouzi and Uzsoy (2014) assume that each information update (i.e., demand quantity update) includes an information gain. This assumption is relaxed in the current setting as temporary forecast biases are explicitly discussed and when a temporary forecast bias occurs, some information updates even imply an information distortion, e.g., based on rationing approaches at the customer as shown in Lee and Whang (2000). The demand forecasts $D_{k,i,j}$ for a final product $k$ at due date $i$ available $j$ periods before delivery are modelled as follows and Table 2 provides a notation table:

$$D_{k,i,j} = \begin{cases} x_{k,i} & \text{for } j > H \\ x_{k,i} + \varepsilon_{k,i,j} & \text{for } j = H \\ D_{k,i,j+1} + \varepsilon_{k,i,j} & \text{for } j < H \end{cases} \quad (3)$$

whereby:
- $x_{k,i}$ is the long-term forecast of the final product $k$ for the due date $i$ (note that we will assume constant long-term forecasts, i.e., $x_{k,i}=x_k$); this is a constant model parameter and set in such a way that the expected order amount $E[D_{k,i,0}]$ is 800 for each product.
- $\varepsilon_{k,i,j}$ is the forecast update term for final product $k$ and due date $i$, which is provided $j$ periods before delivery; this is a stochastic variable drawn from a truncated normal distribution at each update during the simulation run.
- $H$ is the forecast horizon, i.e., forecast updates start $H$ periods before delivery; this is a constant model parameter and set to 10 periods. Note that $\varepsilon_{k,i,0}=0$ is applied for all settings, i.e., no update occurs 0 periods before delivery, and, so, $H=10$ implies 10 forecast updates.

The forecast update terms are defined as:

$$\varepsilon_{k,i,j} \sim N\left(\min, \max, E[\varepsilon_{k,i,j}], \sigma[\varepsilon_{k,i,j}]\right) \\ = N\left(-D_{k,i,j+1}, \left(D_{k,i,j+1} + 2E[\varepsilon_{k,i,j}]\right), E[\varepsilon_{k,i,j}], \sigma[\varepsilon_{k,i,j}]\right) \quad (4)$$

whereby:
- $E[\varepsilon_{k,i,j}]$ is the expected value of forecast updates; note that this value is zero if no forecast bias occurs
- $\sigma[\varepsilon_{k,i,j}]$ is the standard deviation of forecast updates; note that this value determines the forecast information uncertainty.

To model a realistic forecast behaviour, the forecast update terms in Equation (4) follow a normal distribution; however, when creating the random variables in the simulation, the distribution is truncated to avoid negative demand values. This means that—depending on the current demand forecast value and the expected value of the forecast update—the range within which the random variable can be drawn is limited. These limits are represented by the *min* and *max* values in Equation (4). Note that for biased forecasts ($E[\varepsilon_{k,i,j}] \neq 0$), the adaption of the upper bound in the truncated normal distribution is necessary to create a symmetric probability distribution function (PDF) for the stochastic update value. Furthermore, for $E[\varepsilon_{k,i,j}] < 0$ (i.e., underbooking) when $D_{k,i,j+1} \leq -E[\varepsilon_{k,i,j}]$ occurs, the update term has to be set to zero ($\varepsilon_{k,i,j} = 0$) in the simulation for technical reasons (because *min* and *max* do not overlap in this case). To enable different uncertainty settings to be analysed, the expected value and the respective standard deviation of demand forecast updates are parameterized as follows:

$$E[\varepsilon_{k,i,j}] = \beta b_j E[D_{k,i,0}]$$
$$\sigma[\varepsilon_{k,i,j}] = \alpha E[D_{k,i,0}] \quad (5)$$

whereby:
- $\alpha$ is the scenario parameter defining the level of uncertainty without forecast bias; this is a constant model parameter
- $\beta$ is the scenario parameter defining the level of forecast bias; this is a constant model parameter
- $b_j$ defines the forecast update bias *j* periods before delivery, $b_j$ is defined for all $j \in \{1, \ldots, H\}$; $b_j$ are constant model parameters defining the forecast bias scenario.

Note that in our numerical study $E[D_{k,i,0}]=800$ is predefined for all final products and for realizing this, the long-term forecast $x_k=800$ is used for unbiased and temporary biased scenarios, but $x_k = 800\left(1 - \sum_{j=1}^{H} \beta b_j\right)$ has to be applied for permanent biased scenarios.

**Table 2: Notations for forecast evolution model.**

| Notation | Description |
|---|---|
| $n_t$ | Net requirements in period $t$ |
| $s$ | Safety stock |
| $y_{t-1}$ | Stock in period $t-1$ |
| $g_t$ | Gross requirement in period $t$ |
| $r_t$ | Scheduled receipts in period $t$ |
| $\delta$ | Period until net requirements compounded to production order |
| $D_{k,i,j}$ | Demand forecasts for final product $k$, due date $i$, $j$ periods before delivery |
| $x_{k,i}$ | Long-term forecast for product $k$, due date $i$ |
| $\varepsilon_{k,i,j}$ | Forecast update term for product $k$, due date $i$, $j$ periods before delivery |
| $H$ | Forecast horizon |
| $E[\varepsilon_{k,i,j}]$ | Expected value of forecast updates |
| $\sigma[\varepsilon_{k,i,j}]$ | Standard deviation of forecast updates |
| $\alpha$ | Scenario parameter for uncertainty without forecast bias |
| $\beta$ | Scenario parameter for forecast bias |
| $b_j$ | Forecast update bias $j$ periods before delivery |

Table 3 provides possible demand forecast values for the different forecast scenarios applied in this paper. Note that this example shows three possible forecast streams that lead to the same final order amount of 739. This equal final order amount is chosen in the example to simplify the comparison between the forecast behaviours, but it is not related to any inherent model behaviour in the simulation.

Scenario 1 in Table 3 shows a setting where each forecast update leads to an information gain, since all updates are unbiased. Such settings will be applied in Section 4.1 to answer RQ1 and in Section 4.2 to answer RQ2. Following this forecast stream, it can be seen that in Period 29, i.e., 11 periods before the due date 40, the forecast is equal to the long-term forecast, i.e., 800 pcs. The information updates start after that, and, for example, in Period 36, i.e., 4 periods before due date 40, the cumulated updates lead to a forecast of 819 pcs. Scenario 2 shows a setting where the long-term forecast is too low on average and each update leads (on average) to an increase in the demand quantity. Note that in this scenario, each update also implies an information gain. Scenario 3 shows a temporary overbooking situation wherein the long-term forecast on average meets the realized demand; however, customers overbook during the updates, and, hence, overbooking occurs 9, 8, and 7 periods before delivery and the respective overbooked quantities are then reduced again 3, 2, and 1 periods before delivery. Such a temporary overbooking can, for example, be based on rationing aspects to reserve capacity for the uncertain final order quantity as shown in Lee and Whang (2000). In Scenario 3, the biased part of the updates does not imply an information gain.

**Table 3: Example data for three demand forecast scenarios with $\alpha$=0.04, a demand due date of $i$=40, Expected order amount $E[D_{k,i,0}]$=800 and a forecast horizon of $H$=10.**

**Scenario 1, unbiased forecasts with $\beta$=0**

| Current period | 29 | 30 | 31 | 32 | 33 | 34 | 35 | 36 | 37 | 38 | 39 | 40 |
|---|---|---|---|---|---|---|---|---|---|---|---|---|
| Forecast update | $\varepsilon_{40,11}$ | $\varepsilon_{40,10}$ | $\varepsilon_{40,9}$ | $\varepsilon_{40,8}$ | $\varepsilon_{40,7}$ | $\varepsilon_{40,6}$ | $\varepsilon_{40,5}$ | $\varepsilon_{40,4}$ | $\varepsilon_{40,3}$ | $\varepsilon_{40,2}$ | $\varepsilon_{40,1}$ | $\varepsilon_{40,0}$ |
| | 0 | 24 | 32 | -15 | -47 | 123 | 27 | -125 | 56 | -58 | -78 | 0 |
| Forecast value | $D_{40,11}$ | $D_{40,10}$ | $D_{40,9}$ | $D_{40,8}$ | $D_{40,7}$ | $D_{40,6}$ | $D_{40,5}$ | $D_{40,4}$ | $D_{40,3}$ | $D_{40,2}$ | $D_{40,1}$ | $D_{40,0}$ |
| | 800 | 824 | 856 | 841 | 794 | 917 | 944 | 819 | 875 | 817 | 739 | **739** |

**Scenario 2, permanent negative biased forecasts with $\beta$=1**

| Current period | 29 | 30 | 31 | 32 | 33 | 34 | 35 | 36 | 37 | 38 | 39 | 40 |
|---|---|---|---|---|---|---|---|---|---|---|---|---|
| Forecast bias | $b_{11}$ | $b_{10}$ | $b_9$ | $b_8$ | $b_7$ | $b_6$ | $b_5$ | $b_4$ | $b_3$ | $b_2$ | $b_1$ | $b_0$ |
| | 0 | 0.04 | 0.04 | 0.04 | 0.04 | 0.04 | 0.04 | 0.04 | 0.04 | 0.04 | 0.04 | 0 |
| Forecast update | $\varepsilon_{40,11}$ | $\varepsilon_{40,10}$ | $\varepsilon_{40,9}$ | $\varepsilon_{40,8}$ | $\varepsilon_{40,7}$ | $\varepsilon_{40,6}$ | $\varepsilon_{40,5}$ | $\varepsilon_{40,4}$ | $\varepsilon_{40,3}$ | $\varepsilon_{40,2}$ | $\varepsilon_{40,1}$ | $\varepsilon_{40,0}$ |
| | 0 | 56 | 64 | 17 | -15 | 155 | 59 | -93 | 88 | -26 | -46 | 0 |
| Forecast value | $D_{40,11}$ | $D_{40,10}$ | $D_{40,9}$ | $D_{40,8}$ | $D_{40,7}$ | $D_{40,6}$ | $D_{40,5}$ | $D_{40,4}$ | $D_{40,3}$ | $D_{40,2}$ | $D_{40,1}$ | $D_{40,0}$ |
| | 480 | 536 | 600 | 617 | 602 | 757 | 816 | 723 | 811 | 785 | 739 | **739** |

**Scenario 3, temporary negative biased forecasts with $\beta$=1**

| Current period | 29 | 30 | 31 | 32 | 33 | 34 | 35 | 36 | 37 | 38 | 39 | 40 |
|---|---|---|---|---|---|---|---|---|---|---|---|---|
| Forecast bias | $b_{11}$ | $b_{10}$ | $b_9$ | $b_8$ | $b_7$ | $b_6$ | $b_5$ | $b_4$ | $b_3$ | $b_2$ | $b_1$ | $b_0$ |
| | 0 | 0 | 0 | 0.04 | 0.04 | 0.08 | 0 | 0 | -0.08 | -0.04 | -0.04 | 0 |
| Forecast update | $\varepsilon_{40,11}$ | $\varepsilon_{40,10}$ | $\varepsilon_{40,9}$ | $\varepsilon_{40,8}$ | $\varepsilon_{40,7}$ | $\varepsilon_{40,6}$ | $\varepsilon_{40,5}$ | $\varepsilon_{40,4}$ | $\varepsilon_{40,3}$ | $\varepsilon_{40,2}$ | $\varepsilon_{40,1}$ | $\varepsilon_{40,0}$ |
| | 0 | 24 | 32 | 17 | -15 | 187 | 27 | -125 | -8 | -90 | -110 | 0 |
| Forecast value | $D_{40,11}$ | $D_{40,10}$ | $D_{40,9}$ | $D_{40,8}$ | $D_{40,7}$ | $D_{40,6}$ | $D_{40,5}$ | $D_{40,4}$ | $D_{40,3}$ | $D_{40,2}$ | $D_{40,1}$ | $D_{40,0}$ |
| | 800 | 824 | 856 | 873 | 858 | 1045 | 1072 | 947 | 939 | 849 | 739 | **739** |

Final order amount in **bold**; Product index $k$ is omitted since only data for one product is shown

## 4. Simulation Model

The simulation framework models production system behaviour, demand modelling, production planning, and production orders as agents and objects in an AnyLogic discrete-event simulation. These components function as interdependent entities (e.g., functions or Java classes) that interact dynamically throughout the simulation lifecycle. Production orders serve as the central link, evolving as they progress through different processing stages. Each instance tracks key attributes such as item type, lotsize, planned and actual start/end times, shop floor release timestamp, and current status, ensuring accurate propagation of demand updates according to the forecast evolution model. The MRP module integrates the forecasted demand that is based on periods before delivery and the respective forecast updates at each simulation step $t$. These updates define gross requirements for subsequent MRP steps (netting, lot-sizing, backward scheduling, and BOM explosion). Once a production order is generated, its lotsize remains fixed.

The simulation follows an event-driven execution, triggering forecast updates, production planning, and shop floor processes at each *t*, with days as the model time unit. The rolling horizon mechanism shifts the planning window forward at each step. The job shop processing component handles the multi-item, multi-stage production and introduces deterministic processing times and stochastic setup times (lognormal distribution). The inventory management component dynamically adjusts stock levels based on material withdrawals of final products and components.

To enable the large simulation study, the simulation is parallelized across 21 computers (each Intel i5, 32 GB RAM), with independent instances running. The execution follows a three-phase sequence:
1. Pre-simulation setup: Scenario parameters are retrieved from an SQLite database, and the next parameterization is loaded from a PostgreSQL database and injected into the simulation model.
2. Simulation runtime: Demand updates, production planning recalculations, order releases, and shop floor processing are executed.
3. Post-simulation processing: KPIs (e.g., overall costs, inventory cost, backorder costs, and service level) are computed from replication results stored in the PostgreSQL database.

By integrating these components, the simulation framework efficiently captures the interactions between demand forecasting, production planning, and shop floor execution. The rolling horizon adjustments, production orders, and demand updates realistically model MRP system behaviour under forecast uncertainty.

## 5. Numerical Study

To address the research questions outlined in Section 1, a simulation model was developed, and a numerical study was conducted. This section describes the experimental setup used in the numerical study, defining the simulation scenarios, key parameters, and performance indicators to evaluate planning strategies under forecast uncertainty and bias.

### *5.1. Scenario Definition and Tested Planning Parameters*

The production system introduced in Section 3.1 is simulated with the forecast uncertainty parameter $\alpha \in \{0.02, 0.04, \ldots, 0.12\}$, level of forecast bias $\beta \in \{0, 1\}$, forecast horizon $H=10$, and the following four update bias behaviours defined in Table 4. Note that according to the modelling in Equation (5), different forecast biases could also be studied by changing $\beta$.

**Table 4: Forecast bias scenarios.**

|  | $b_{10}$ | $b_9$ | $b_8$ | $b_7$ | $b_6$ | $b_5$ | $b_4$ | $b_3$ | $b_2$ | $b_1$ |
|---|---|---|---|---|---|---|---|---|---|---|
| Temporary overbooking | 0 | 0 | 0.04 | 0.04 | 0.08 | 0 | 0 | -0.08 | -0.04 | -0.04 |
| Temporary underbooking | 0 | 0 | -0.04 | -0.04 | -0.08 | 0 | 0 | 0.08 | 0.04 | 0.04 |
| Permanent overbooking | -0.04 | -0.04 | -0.04 | -0.04 | -0.04 | -0.04 | -0.04 | -0.04 | -0.04 | -0.04 |
| Permanent underbooking | 0.04 | 0.04 | 0.04 | 0.04 | 0.04 | 0.04 | 0.04 | 0.04 | 0.04 | 0.04 |

The range in $\alpha$ values is chosen to investigate the effect of increasing forecast uncertainty, while $\beta=1$ is applied to investigate the influence of forecast bias. The $\alpha$, $\beta$, and $b_j$ values have been identified in simulation pre-tests to be in a reasonable range related to the introduced production system. Overall, this leads to 21 unbiased test instances, i.e., 7 $\alpha$-values tested for 3 utilization values, and 84 biased test instances, i.e., a full factorial design of 7 $\alpha$-values, 1 $\beta$-value (as $\beta=0$ is unbiased), 4 forecast bias scenarios, and 3 utilization values.

In this paper, we focus on optimizing the MRP parameters lotsize, planned lead time, and safety stock related to forecast uncertainty effects. Since typical production systems are multi-stage, but the main effect of forecast uncertainty is on the production orders of final products, a two-stage production system is simulated. However, the planning parameters are only optimized for the products.

Some preliminary simulation experiments were conducted to optimize the lotsize and planned lead time for components, assuming zero safety stock for components. These preliminary experiments applied a reduced set of final product lotsizes and different α values for $\beta=0$. The results, based on preliminary studies with the evaluated production system model, indicate that for components, a planned lead time of 3 is appropriate, and two different lotsizes, i.e., FOQ 800 and FOQ 1600, must be tested. To identify the optimal planning parameters for both *MRP standard* and *MRP safety stock exploitation heuristic*, a solution space enumeration, i.e., full factorial design, is conducted for each test instance. The parameter range is derived from the preliminary studies and is limited by the enumeration setting, covering:

- Planned lead time PLT $\in \{1,2,3,4,6,8\}$,
- Safety stock factor SST $\in \{0,0.2,0.4,0.6,0.8,1,1.5,2\}$,
- FOP $\in \{1,2,5,6,9\}$, and FOQ $\in \{200,400,800,1200,1600\}$.

Either FOP or FOQ lot-sizing is applied. In combination with the two component lotsizes, this results in 960 planning parameter sets per test instance. These 960 parameter settings are tested for both *MRP standard* and the *MRP safety stock exploitation heuristic*. To create valid results, 20 replications for the same parameter set are conducted in the simulation experiment and the simulation run time is 400 periods with warm-up for 40 periods. For the numerical study, 4,032,000 simulation runs, each lasting 8-10 seconds on a single processor core, were performed.

*5.2. Performance Measurement*

The performance of the production system, i.e., the overall costs, are evaluated as the sum of WIP costs (0.5 Cost Units (CU)/period), FGI costs (1 CU/period), and backorder costs (19 CU/period). All costs are measured per piece of a component or final product. Since final products have higher capital binding costs than WIP, FGI holding costs are twice as high as WIP costs. Additionally, backorder costs are set to 19 CU/period, based on a target service level of 95%, calculated using the formula: $service\ level = 1 - (FGI\ costs)/(FGI\ costs + backorder\ costs)$, refer to Axsäter (2015) for more details. Note that all demands which cannot be fulfilled are backordered and demands can only be fulfilled fully. Thus, in the case the required demand of the final products is not on stock at the due date, the whole demand is backordered until it can be fully fulfilled. For certain in-depth analysis of the planning method behaviour also the service level is evaluated.

Since each scenario, defined by the forecast uncertainty parameter $\alpha$ and the forecast bias parameter $\beta$, represents a unique combination of forecast uncertainty and bias levels, the lowest overall costs are determined and compared independently. A direct comparison using the same parameters would be misleading, as the MRP standard relies on fixed safety stock settings, while the safety stock exploitation heuristic dynamically adjusts stock levels based on forecast updates. Thus, the parameter set yielding the lowest overall costs for each scenario is reported. This approach allows the heuristic to fully leverage its advantages, such as improved forecast responsiveness and reduced MRP nervousness, without being limited by shared parameters. In contrast, the MRP standard requires higher fixed safety stocks to buffer demand uncertainty, making independent optimization necessary for a fair comparison.

## 6. Numerical Results

The results are presented in three parts. First, the performance of the *MRP standard* is assessed, providing an in-depth analysis of parameterization and production performance under varying levels of forecast uncertainty at low utilization and no forecast bias. Second, the *MRP safety stock exploitation heuristic* is evaluated for the identical scenario. Finally, a comparative performance evaluation highlights cost reduction potential and parameterization differences between the *MRP standard* and the *MRP safety stock exploitation heuristic* across unbiased, permanent biased, and temporary biased forecast scenarios. Each subsection concludes with a summary of key insights derived from the analysis.

*6.1. MRP Standard*

The impact of different parameter settings on overall costs and service levels is examined for *MRP standard*. This is followed by an in-depth analysis of parameterization and production system performance under varying levels of forecast uncertainty at low utilization and no forecast bias.

### 6.1.1. Analysis of Planning Parameter Effects

To foster the understanding of how different planning parameter settings influence the overall costs and the resulting service level, a detailed analysis for the low utilization setting with $\alpha$=0.04 and $\beta$=0 is presented here. For the lotsizing policies FOP and FOQ, detailed results for selected lotsizing and planned lead time parameters as well as the best performing parameter combinations with respect to predefined safety stock levels are shown in Figure 2a–2f. The safety stock factor is related to the expected demand per due date E[$D_{k,i,0}$]=800, i.e., a safety stock factor of 0.5 means a safety stock level of 400 pcs.

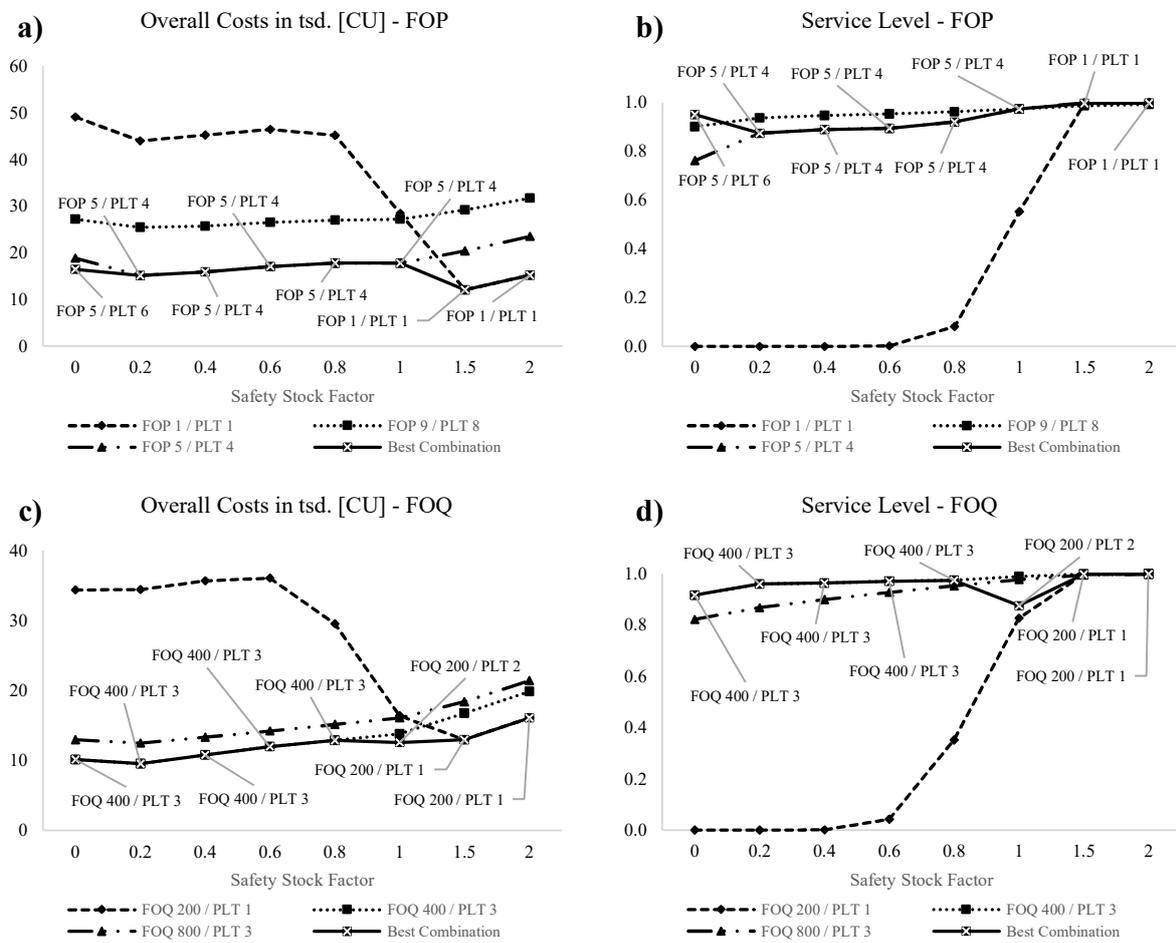

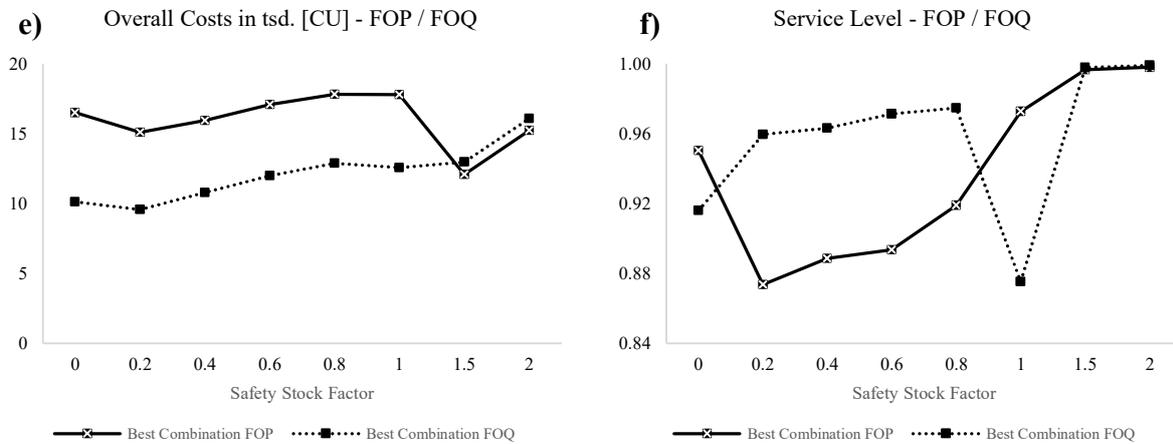

**Figure 2: Detailed cost and service level analysis with respect to safety stock level for MRP standard.**

The results of Figure 2 (a-f) support the model validity and show that higher safety stocks lead to a higher service level and that too-high levels of safety stocks have a negative impact on overall costs, based on high inventory costs. Comparing the best overall costs of FOP and FOQ in Figure 2e shows that FOQ significantly outperforms FOP from 0 to 1 as safety stock factor, which is related to the high amount of short-term orders created with FOP in MRP standard, when forecast updates occur. An examination of the optimal lotsizes and planned lead times of FOP in Figure 2a and 2b shows that for low safety stocks, high lotsizes and high planned lead times are applied, and for high safety stocks (1, 1.5, and 2), FOP=1 and PLT=1 are applied, which means a make-to-stock-system is mimicked where a stock withdrawal leads to new production orders. The FOQ Figures 2c and 2d show a similar behaviour for high safety stocks. However, the lowest costs are reached with lower safety stocks (safety stock=0.2) and planned lead times (PLT=3), which also include future demand information.

### 6.1.2. *Optimal Planning Parameters at Low Utilization*

The influence of forecast uncertainty on optimal planning parameters for the MRP standard is discussed in this subsection for low utilization and no forecast bias $\beta=0$. Table 5 shows the optimal planning parameters for FOP and FOQ related to the forecast uncertainty $\alpha$. The column lead time final prod reports the production order lead time of final products along with its standard deviation (±), which reflects the stochastic nature of realized lead times. The standard deviation is not intended to indicate the statistical significance of lead times for different $\alpha$ values but rather to illustrate how lead time variability increases as forecast uncertainty rises. The number of production orders (*# of prod.*) shows the average number of production orders issued for final products. An intuitive result is that higher forecast uncertainty leads to higher overall costs, which is in line with Altendorfer et al. (2016) and Enns (2002). In detail, the results illustrate that FOQ consistently outperforms FOP lotsizing related to costs. An interesting result is that PLT=1 and FOP=1 is optimal for all $\alpha$-settings when FOP is applied. This means that for FOP lotsizing, MRP tries to mimic a make-to-stock system without benefiting from demand forecast information. For FOQ, the optimal planned lead time for low to medium forecast uncertainty is significantly higher than the realized lead time of final products production orders, i.e., when FOQ is applied, MRP benefits from demand forecasts. We conjecture that this is the reason why FOQ outperforms FOP in all instances. However, for higher forecast uncertainties ($\alpha \geq 0.1$), PLT=1 and SST=1.5 are optimal for FOQ. This trend is less clear when considering only the service level, as FOP achieves higher service levels at lower forecast uncertainties ($\alpha \leq 0.08$). However, focusing solely on service level can be misleading. For example, at $\alpha$=0.08, although FOP achieves a better service level, it results in higher backorder costs. This discrepancy arises because service level calculations do not account for the order quantity or the associated tardiness duration, both of which directly impact overall cost performance. In addition to the reported test instances, $\alpha$=0 has also been tested, and the lowest costs there are CU 4.158, which are equal for FOQ and FOP (since both lead to a constant lotsize of 800 pcs). This shows that information uncertainty leads to significant additional costs, even if the uncertainty is very low. Comparing the standard deviation of the production lead time between FOP and FOQ shows that, specifically for higher $\alpha$-values, the FOP policy has significantly higher uncertainties in production lead time which implies higher system nervousness. From a managerial point of view, this leads to the following insight:

***Insight 1****: When applying MRP standard without manual planner interaction, the FOP lotsizing policy leads to additional MRP nervousness and MRP cannot benefit from uncertain forecast information; hence, the FOQ lotsizing policy should be favoured.*

**Table 5: MRP standard – Optimal planning parameters for unbiased forecasts and low utilization with respect to FOP and FOQ lotsize policy.**

| α | FOP Lotsize | | | | | | | | FOQ Lotsize | | | | | | | |
|---|---|---|---|---|---|---|---|---|---|---|---|---|---|---|---|---|
| | SST | PLT | Lotsize | Overall Costs [CU] | # of prod. | Backorder Costs [CU] | Service Level | Leadtime Final Prod | SST | PLT | Lotsize | Overall Costs [CU] | # of prod. | Backorder Costs [CU] | Service Level | Leadtime Final Prod |
| 0.02 | 1.5 | 1 | 1 | 12,111 | 1,070 | 0 | 100% | 1.85±0.19 | 0.2 | 3 | 200 | 7,830 | 837 | 117 | 99% | 1.95±0.14 |
| 0.04 | 1.5 | 1 | 1 | 12,068 | 1,073 | 141 | 100% | 2.13±0.48 | 0.2 | 3 | 400 | 9,548 | 810 | 411 | 96% | 2.13±0.26 |
| 0.06 | 1.5 | 1 | 1 | 12,847 | 1,076 | 451 | 98% | 2.43±0.62 | 0.2 | 3 | 400 | 10,254 | 848 | 1,126 | 91% | 2.29±0.38 |
| 0.08 | 1.5 | 1 | 1 | 15,153 | 1,077 | 2,865 | 92% | 2.8±1.07 | 0.2 | 4 | 400 | 12,441 | 931 | 1,843 | 91% | 2.78±0.98 |
| 0.10 | 2.0 | 1 | 1 | 16,808 | 1,077 | 1,674 | 94% | 3.14±1.53 | 1.5 | 1 | 200 | 14,803 | 830 | 1,448 | 97% | 2.37±0.51 |
| 0.12 | 2.0 | 1 | 1 | 18,365 | 1,079 | 3,356 | 91% | 3.46±1.92 | 1.5 | 1 | 400 | 15,992 | 759 | 1,971 | 97% | 2.61±0.72 |

## 6.2. MRP Extended by Safety Stock Exploitation Heuristic

An in-depth analysis of parameterization and production system performance under varying levels of forecast uncertainty at low utilization and no forecast bias is conducted for the *MRP safety stock exploitation heuristic*, following the same approach as for the *MRP standard*. Table 6, shows the optimal planning parameters for FOP and FOQ for low utilization and different forecast uncertainty levels. A comparison of the overall costs between Table 5 and Table 6 shows that the *MRP safety stock exploitation heuristic* significantly outperforms MRP standard for FOP and FOQ. Furthermore, the results show that in this setting for FOP, the optimal planning parameters depend on the forecast uncertainty; also, for high uncertainty, the system still benefits from uncertain information. A comparison of FOP and FOQ shows that for very low uncertainty, FOP performs better, but starting with $\alpha=0.04$, FOQ leads to lower costs. This shows that the advantage of FOP to produce exactly the demanded amounts must be traded off with the advantage of FOQ to create some buffer because of the higher production lotsizes compared to the forecasted demand. Furthermore, FOQ always leads to a lower amount of production orders in the optimal setting, which implies that FOP generates (even in the optimal setting and applying the *MRP safety stock exploitation heuristic*) more short-term orders with low lotsizes to react on forecast updates. Upon analysing the results presented in Table 5 and Table 6, it becomes evident that our *safety stock exploitation heuristic*, when optimally parameterized, not only reduces the production lead times of final products but also significantly minimizes their fluctuation (as indicated by the +/- term, representing standard deviation). This effectively demonstrates a reduction in MRP nervousness. Hence, in comparison to the findings of Sridharan and Lawrence LaForge (1989), our *safety stock exploitation heuristic* lead to a marked decrease in MRP nervousness (schedule instability), showcasing a distinct advantage over their approach.

As regards the lotsizing parameter, an interesting finding is that the application of the *MRP safety stock exploitation heuristic* leads to FOP 1 or 2 and FOQ 200 as best parameters, which both imply low production lotsizes. Note that for FOQ, the parameter 200 means that production orders are very near the forecasted demand, but always a multiple of 200 is applied when an order is issued (see (Hopp and Spearman 2011) for details on FOQ policy). This lot-sizing behaviour directly impacts service levels, where FOP initially maintains higher fulfilment rates but struggles as forecast uncertainty increases, while FOQ stabilizes at medium α values and outperforms FOP under high uncertainty.

**Table 6: MRP extended – Optimal planning parameters for unbiased forecasts and low utilization.**

| | | | | FOP Lotsize | | | | | | | FOQ Lotsize | | | | | |
|---|---|---|---|---|---|---|---|---|---|---|---|---|---|---|---|---|
| α | SST | PLT | FOP | Overall Costs [CU] | # of prod. | Backorder Costs [CU] | Service Level | Leadtime Final Prod | SST | PLT | Lotsize | Overall Costs [CU] | # of prod. | Backorder Costs [CU] | Service Level | Leadtime Final Prod |
| 0.02 | 0.2 | 2 | 1 | 6,149 | 739 | 872 | 94% | 1.82±0.02 | 0.2 | 2 | 200 | 7,466 | 720 | 1,175 | 80% | 1.88±0.1 |
| 0.04 | 0.6 | 2 | 1 | 8,315 | 728 | 1,267 | 79% | 1.88±0.06 | 0.2 | 3 | 200 | 8,013 | 721 | 285 | 99% | 1.96±0.2 |
| 0.06 | 0.6 | 3 | 2 | 10,450 | 745 | 1,225 | 96% | 2±0.15 | 0.2 | 3 | 200 | 8,695 | 728 | 325 | 99% | 2.02±0.12 |
| 0.08 | 0.6 | 3 | 2 | 12,166 | 761 | 2,871 | 90% | 2.18±0.33 | 0.4 | 3 | 200 | 9,851 | 720 | 322 | 96% | 2.14±0.2 |
| 0.10 | 1.0 | 3 | 2 | 14,846 | 736 | 3,795 | 89% | 2.36±0.75 | 0.4 | 3 | 200 | 10,830 | 721 | 1,384 | 88% | 2.3±0.34 |
| 0.12 | 0.8 | 3 | 2 | 18,262 | 758 | 7,869 | 78% | 2.66±1.34 | 0.6 | 4 | 200 | 12,655 | 713 | 510 | 97% | 2.5±0.58 |

### *6.3. Comparative Performance Evaluation (MRP Standard vs. Extended)*

A comparative performance evaluation of the *MRP standard* and the *MRP safety stock exploitation heuristic* is conducted, focusing on cost reduction potential and parameterization differences across unbiased, permanent biased, and temporary biased forecast scenarios. The planning parameters for both methods are independently optimized and may not necessarily be identical, and the lot-sizing policy yielding the lowest overall costs for each specific scenario is reported. The cost reduction potential, expressed as a percentage, is calculated by comparing the overall costs of the *MRP safety stock exploitation heuristic* with those of the *MRP standard*. To ensure a comprehensive assessment, confidence intervals are computed for α=0.05 and α=0.01, with (*) marking significance at α=0.05 and (**) at α=0.01.

#### 6.3.1. Unbiased Forecast

Table 7 presents a comprehensive comparison of all 21 test instances for unbiased forecasts, evaluating the performance of the *MRP standard* and the *MRP safety stock exploitation heuristic*, with the lot-sizing policy yielding the lowest overall costs consistently reported. Overall, the results of Table 7 show that the *MRP safety stock exploitation heuristic* leads to a significant cost reductions compared to the MRP standard across all tested scenarios. Most of these cost reductions are highly statistically significant (**), providing strong evidence that the heuristic consistently outperforms the standard approach. However, in scenarios with both high utilization and high forecast uncertainty ($\alpha \geq 0.10$), statistical significance is not observed, indicating that performance differences are less pronounced under these conditions.

As seen in Table 7, higher shop load, in general, leads to higher overall costs, which is an intuitive result and supports model validity. For the MRP standard, higher forecast uncertainty leads to higher optimal lotsizes, which implies that more buffer is included in the production orders. For the MRP standard, the optimal planned lead time and safety stocks show that for higher shop loads and higher forecast uncertainty, the production system does not benefit from uncertain demand forecasts. These observations closely align with the "Buffering Law" from Hopp and Spearman (2011), where failing to invest in reducing uncertainty inevitably results in performance decrease elsewhere, i.e., high inventory which leads to longer production lead times based on Little's Law (Little 1961).

The results for the *MRP safety stock exploitation heuristic* show that the application of the heuristic enables the production planning to include uncertain demand forecasts and benefit from the uncertain demand forecast information. Only for high shop load and $\alpha \geq 0.08$, the *MRP safety stock exploitation heuristic* also applies high safety stocks and low planned lead times, i.e., no benefit is derived from uncertain demand forecast information. Also, in these test instances, the cost reduction potential of the *MRP safety stock exploitation heuristic* is very low. From a managerial point of view, the following insights are gained:

***Insight 2****: The application of the MRP safety stock exploitation heuristic leads to significant cost reduction potentials and enables the production planning to benefit from uncertain demand forecasts without manual planner intervention.*

***Insight 3****: When uncertain (and unbiased) demand forecast information with updates occurs, the FOQ lotsizing policy leads for almost all cases (except very low uncertainty levels) to a better performance than FOP.*

***Insight 4****: An increase in forecast uncertainty must be buffered by higher safety stocks and higher planned lead times; however, it always implies significantly higher costs.*

**Table 7: Optimal overall costs and planning parameter comparison for unbiased forecasts with respect to MRP standard and MRP safety stock exploitation heuristic (MRP extended).**

| | Low Utilization | | | | | | | | | | |
|---|---|---|---|---|---|---|---|---|---|---|---|
| | MRP standard | | | | | MRP extended | | | | | Cost red. MRP stand. |
| α | SST | PLT | Lotsize | Overall Costs [CU] | Service Level | SST | PLT | Lotsize | Overall Costs [CU] | Service Level | |
| 0.02 | 0.2 | 3 | FOQ 200 | 7,830 | 99% | 0.2 | 2 | FOP 1 | 6,149 | 94% | -21% ** |
| 0.04 | 0.2 | 3 | FOQ 400 | 9,548 | 96% | 0.2 | 3 | FOQ 200 | 8,013 | 99% | -16% ** |
| 0.06 | 0.2 | 3 | FOQ 400 | 10,254 | 91% | 0.2 | 3 | FOQ 200 | 8,695 | 99% | -15% ** |
| 0.08 | 0.2 | 4 | FOQ 400 | 12,441 | 91% | 0.4 | 3 | FOQ 200 | 9,851 | 96% | -21% ** |
| 0.10 | 1.5 | 1 | FOQ 200 | 14,803 | 97% | 0.4 | 3 | FOQ 200 | 10,830 | 88% | -27% ** |
| 0.12 | 1.5 | 1 | FOQ 400 | 15,992 | 97% | 0.6 | 4 | FOQ 200 | 12,655 | 97% | -21% ** |

| | Medium Utilization | | | | | | | | | | |
|---|---|---|---|---|---|---|---|---|---|---|---|
| | MRP standard | | | | | MRP extended | | | | | Cost red. MRP stand. |
| α | SST | PLT | Lotsize | Overall Costs [CU] | Service Level | SST | PLT | Lotsize | Overall Costs [CU] | Service Level | |
| 0.02 | 0.0 | 3 | FOQ 400 | 8,967 | 90% | 0.4 | 2 | FOP 1 | 6,967 | 80% | -22% ** |
| 0.04 | 1.5 | 1 | FOQ 200 | 12,746 | 100% | 0.2 | 3 | FOQ 200 | 7,967 | 99% | -37% ** |
| 0.06 | 1.5 | 1 | FOQ 200 | 13,172 | 98% | 0.2 | 3 | FOQ 200 | 9,320 | 92% | -29% ** |
| 0.08 | 1.5 | 1 | FOQ 400 | 14,004 | 99% | 0.4 | 4 | FOQ 200 | 11,080 | 98% | -21% ** |
| 0.10 | 1.5 | 1 | FOQ 400 | 15,684 | 94% | 0.6 | 4 | FOQ 200 | 13,104 | 94% | -16% ** |
| 0.12 | 1.5 | 1 | FOQ 800 | 17,090 | 94% | 0.6 | 4 | FOQ 200 | 15,077 | 87% | -12% * |

| | High Utilization | | | | | | | | | | |
|---|---|---|---|---|---|---|---|---|---|---|---|
| | MRP standard | | | | | MRP extended | | | | | Cost red. MRP stand. |
| α | SST | PLT | Lotsize | Overall Costs [CU] | Service Level | SST | PLT | Lotsize | Overall Costs [CU] | Service Level | |
| 0.02 | 1.0 | 1 | FOQ 400 | 12,469 | 96% | 0.2 | 3 | FOQ 200 | 7,630 | 100% | -39% ** |
| 0.04 | 1.5 | 1 | FOQ 200 | 13,044 | 98% | 0.2 | 3 | FOQ 200 | 8,439 | 92% | -35% ** |
| 0.06 | 1.5 | 1 | FOQ 400 | 14,179 | 94% | 0.4 | 4 | FOQ 200 | 11,344 | 95% | -20% ** |
| 0.08 | 1.5 | 1 | FOQ 800 | 15,513 | 96% | 1.5 | 1 | FOQ 200 | 13,878 | 94% | -11% ** |
| 0.10 | 1.5 | 1 | FOQ 800 | 17,380 | 92% | 1.5 | 1 | FOQ 800 | 15,888 | 94% | -9% |
| 0.12 | 2.0 | 1 | FOQ 800 | 19,098 | 96% | 2.0 | 2 | FOQ 200 | 17,639 | 95% | -8% * |

### 6.3.2. Permanent forecast bias

To identify the effect of biased forecasts on MRP performance, this section focuses on permanent forecast biases. In the underbooking case, the long-term forecast is significantly lower than the expected final order amount, i.e., customers underestimate their demand, and each update leads, on average, to a higher forecast value. For overbooking, customers overestimate their demand, and each update leads to a reduction in the forecasted demand. Despite this, each update still provides an information gain. In the overbooking case, production orders are, on average, too high, and in the rolling horizon setting, an undefined buffer stock arises. This buffer stock increases inventory costs but can absorb short-term demand increases from unsystematic uncertainty. In the underbooking case, too few production orders are created, and with short-term demand increases from unsystematic uncertainty, demand rises systematically after production orders are created.

**Table 8: Optimal overall costs and planning parameter comparison for permanent forecast bias.**

| | | | | Low Utilization MRP extended | | | | | | | | | |
|---|---|---|---|---|---|---|---|---|---|---|---|---|---|
| | | | | Permanent Overbooking | | | | | | Permanent Underbooking | | | |
| $\beta$ | $\alpha$ | SST | PLT | Lotsize | Overall Costs [CU] | Service Level | Cost red. MRP stand. | | SST | PLT | Lotsize | Overall Costs [CU] | Service Level | Cost red. MRP stand. |
| 1 | 0.02 | 0.0 | 2 | FOP 1 | 5,396 | 97% | 0% | | 0.2 | 2 | FOQ 200 | 6,891 | 75% | -26% ** |
| 1 | 0.04 | 0.2 | 2 | FOP 1 | 7,705 | 78% | -3% ** | | 0.4 | 3 | FOQ 200 | 8,084 | 99% | -23% ** |
| 1 | 0.06 | 0.0 | 3 | FOQ 200 | 8,644 | 96% | 0% | | 0.6 | 3 | FOQ 200 | 8,603 | 99% | -29% ** |
| 1 | 0.08 | 0.2 | 3 | FOP 2 | 9,206 | 97% | -13% ** | | 0.8 | 3 | FOQ 200 | 9,650 | 95% | -29% ** |
| 1 | 0.10 | 0.2 | 3 | FOQ 200 | 10,775 | 89% | -18% ** | | 0.8 | 3 | FOQ 200 | 11,288 | 86% | -33% ** |
| 1 | 0.12 | 0.2 | 4 | FOQ 200 | 12,722 | 95% | -14% ** | | 1.0 | 3 | FOQ 200 | 14,022 | 78% | -25% ** |

| | | | | Medium Utilization MRP extended | | | | | | | | | |
|---|---|---|---|---|---|---|---|---|---|---|---|---|---|
| | | | | Permanent Overbooking | | | | | | Permanent Underbooking | | | |
| $\beta$ | $\alpha$ | SST | PLT | Lotsize | Overall Costs [CU] | Service Level | Cost red. MRP stand. | | SST | PLT | Lotsize | Overall Costs [CU] | Service Level | Cost red. MRP stand. |
| 1 | 0.02 | 0.0 | 2 | FOP 1 | 6,063 | 71% | 0% | | 0.4 | 3 | FOQ 200 | 6,982 | 99% | -29% ** |
| 1 | 0.04 | 0.0 | 3 | FOQ 200 | 7,880 | 98% | 0% | | 0.4 | 3 | FOQ 200 | 8,074 | 97% | -30% ** |
| 1 | 0.06 | 0.2 | 3 | FOP 1 | 8,816 | 97% | -16% ** | | 0.6 | 3 | FOQ 200 | 8,957 | 92% | -33% ** |
| 1 | 0.08 | 0.2 | 3 | FOP 2 | 10,912 | 85% | -21% ** | | 1.0 | 4 | FOQ 200 | 11,172 | 98% | -31% ** |
| 1 | 0.10 | 0.4 | 4 | FOP 1 | 12,884 | 95% | -18% ** | | 1.0 | 4 | FOQ 200 | 13,349 | 91% | -25% ** |
| 1 | 0.12 | 0.4 | 4 | FOQ 200 | 14,754 | 89% | -15% ** | | 1.5 | 4 | FOQ 400 | 16,357 | 87% | -21% ** |

| | | | | High Utilization MRP extended | | | | | | | | | |
|---|---|---|---|---|---|---|---|---|---|---|---|---|---|
| | | | | Permanent Overbooking | | | | | | Permanent Underbooking | | | |
| $\beta$ | $\alpha$ | SST | PLT | Lotsize | Overall Costs [CU] | Service Level | Cost red. MRP stand. | | SST | PLT | Lotsize | Overall Costs [CU] | Service Level | Cost red. MRP stand. |
| 1 | 0.02 | 0.0 | 3 | FOP 1 | 6,968 | 99% | 0% | | 0.4 | 3 | FOQ 200 | 6,980 | 98% | -31% ** |
| 1 | 0.04 | 0.2 | 3 | FOP 1 | 8,704 | 97% | -9% ** | | 0.6 | 3 | FOQ 200 | 8,796 | 92% | -25% ** |
| 1 | 0.06 | 0.2 | 4 | FOP 1 | 10,556 | 96% | -17% ** | | 0.8 | 4 | FOQ 200 | 10,916 | 94% | -26% ** |
| 1 | 0.08 | 0.4 | 4 | FOP 2 | 14,349 | 88% | -5% | | 1.0 | 4 | FOQ 200 | 14,822 | 83% | -18% |
| 1 | 0.10 | 1.0 | 1 | FOQ 800 | 16,326 | 93% | -4% | | 1.5 | 4 | FOP 5 | 17,182 | 93% | -20% ** |
| 1 | 0.12 | 1.0 | 2 | FOQ 800 | 17,669 | 95% | -5% | | 2.0 | 4 | FOP 5 | 18,405 | 95% | -32% ** |

In general, the results in Table 8 show that cost reductions are consistently highly significant (**) in the permanent underbooking scenarios, while in overbooking, for low $\alpha$ values no cost improvement is gained and for some of the high utilization scenarios the cost improvement is not significant (no stars). Both overbooking and underbooking show service level fluctuations across different $\alpha$ values rather than a consistent trend. This further reinforces that focusing solely on service level can be misleading, as it excludes order quantity and tardiness duration, both of which directly impact overall cost performance.

Moreover, related to the lowest costs and planning parameters for *MRP safety stock exploitation heuristic*, reported in Table 8, reveal that optimal costs are generally higher in the underbooking case than in the overbooking case. This cost difference is primarily due to the higher backorder cost factor compared to the inventory cost factor, combined with the rolling horizon nature of MRP-based production planning. This leads to an asymmetry in stock creation: in overbooking, additional stock is more likely to be generated, while in underbooking, backorders are more frequent. Although not directly visible in the tables, this effect stems from forecasted order amounts influencing production decisions. In underbooking, underestimated forecasts lead to insufficient production orders, increasing the risk of backorders and cost penalties. This general finding is in line with Enns (2002), where also overestimated forecasts lead to a better performance than underestimated ones. However, in Enns (2002), no forecast update behaviour is modelled and the MRP standard is applied. The results of permanent overbooking, when analysed in detail, show that the costs for low $\alpha$ values are significantly lower than in the unbiased setting, independent of the utilization. Furthermore, the costs with overbooking are lower than in the unbiased situation, when utilization is low to medium; only at high utilizations are there some instances where the unbiased setting leads to lower costs. In addition, the safety stock at the overbooking case, too, is lower than in the unbiased case.

The results for underbooking show that in most instances, the costs are higher than in the unbiased case, and the safety stocks are higher as well. This supports the observations of Sridharan and Lawrence LaForge (1989), who noted the moderate hedging of the MPS, which overbooking leads to, is favourable. However, they used safety stock, instead of forecast evolution for hedging the MPS. Furthermore, only FOQ is the best lot sizing policy there, except in the case where utilization and forecast uncertainty are high. Comparing this result with the overbooking scenario leads to the interesting finding that low forecast uncertainty and the undefined buffer from overbooking provides the possibility to apply FOP and create production orders in a quantity that better fits the demand. A general finding is that in the permanent biased settings, *MRP safety stock exploitation heuristic* leads to a better (or, in a few cases, equal) cost performance than the MRP standard. In the overbooking case, *MRP safety stock exploitation heuristic* has a significantly higher cost reduction potential compared to the MRP standard than in the underbooking case. The following insights can be noted.

***Insight 5****: A systematic overestimation of long-term demand and the respective step-wise reduction by periodic updates, i.e., forecast evolution with permanent overbooking, has a positive effect on overall cost for low to medium utilizations and leads to significantly lower safety stocks than in the unbiased case. For better forecasts, i.e., lower uncertainty, FOP should be applied and FOQ otherwise.*

***Insight 6****: A comparison of permanent overbooking and underbooking shows that an underbooking behaviour of customers is much more expensive for a manufacturing company than an overbooking behaviour, i.e., for customers this means that for uncertain long-term demand it is better to overestimate demands than to underestimate them.*

From a practical point of view, it is an interesting finding that such an overestimation led to a cost reduction of 3% on average over all instances compared to the unbiased setting (comparing the sum of unbiased and permanent biased lowest costs); i.e., even the manufacturing company benefits from overbooking.

### 6.3.3. Temporary forecast bias

To provide a comprehensive picture of systematic forecasting behaviour, the results for temporary overbooking and underbooking are shown in this section. Temporary means that the long-term demand fits the average order amount, but the forecast updates include either a temporary demand increase or a decrease in addition to the unsystematic error. Here, this means that not each update includes an information gain, but some updates also distort the information provided. The increase can be based on rationing objectives of the customer while the decrease might be a systematic effect in the customers planning system.

The respective results in Table 9 for *MRP safety stock exploitation heuristic* show a similar behaviour than the ones in Table 8. In general, statistical significance remains strong in the temporary underbooking scenarios (**), while in temporary overbooking, a similar result as in the permanent overbooking case is observed, with no improvement for low $\alpha$ values and some statistically not significant scenarios at high utilization. An examination of the optimal planning parameters shows that in this temporary overbooking and underbooking case, the make-to-stock setting with low planned lead time and high safety stocks is used for more test instances than in the permanent forecast bias case, specifically for medium to high unsystematic uncertainties. This might be related to the additional disturbance created by having firstly an increase/decrease of forecasts and then a decrease/increase respectively, and the fact that not all forecast updates include an information gain. Wijngaard (2004) supports this observation, noting the limited improvement from advanced demand information in tight capacity situations. However, his study focused on a single-stage, single-item production system without integrating demand information updates. This observation leads to the following insight:

***Insight 7:*** *For production systems with high congestion, temporary biased forecasts, and high uncertainty in forecasts, the MRP changes to a make-to-stock planning system, ignoring future forecast information, in which only stock withdrawals are re-produced.*

**Table 9: Comparison of optimal overall costs and planning parameters for temporary forecast bias.**

| | | Low Utilization MRP extended | | | | | | | | | | | |
|---|---|---|---|---|---|---|---|---|---|---|---|---|---|
| | | Temporary Overbooking | | | | | | Temporary Underbooking | | | | | |
| β | α | SST | PLT | Lotsize | Overall Costs [CU] | Service Level | Cost red. MRP stand. | SST | PLT | Lotsize | Overall Costs [CU] | Service Level | Cost red. MRP stand. |
| 1 | 0.02 | 0.0 | 2 | FOP 1 | 5,396 | 97% | 0% | 0.2 | 2 | FOQ 200 | 7,154 | 75% | -27% ** |
| 1 | 0.04 | 0.2 | 2 | FOP 1 | 7,185 | 78% | -7% ** | 0.4 | 3 | FOQ 200 | 8,366 | 99% | -35% ** |
| 1 | 0.06 | 0.0 | 3 | FOQ 200 | 8,276 | 97% | 0% | 0.4 | 3 | FOQ 200 | 8,685 | 99% | -34% ** |
| 1 | 0.08 | 0.2 | 3 | FOP 2 | 9,220 | 96% | -11% ** | 0.4 | 3 | FOQ 200 | 9,853 | 93% | -28% ** |
| 1 | 0.10 | 0.4 | 3 | FOP 1 | 10,824 | 91% | -15% ** | 0.6 | 3 | FOQ 200 | 11,282 | 88% | -24% ** |
| 1 | 0.12 | 0.4 | 4 | FOQ 200 | 12,594 | 97% | -16% ** | 0.8 | 4 | FOQ 200 | 12,874 | 97% | -22% ** |

| | | Medium Utilization MRP extended | | | | | | | | | | | |
|---|---|---|---|---|---|---|---|---|---|---|---|---|---|
| | | Temporary Overbooking | | | | | | Temporary Underbooking | | | | | |
| β | α | SST | PLT | Lotsize | Overall Costs [CU] | Service Level | Cost red. MRP stand. | SST | PLT | Lotsize | Overall Costs [CU] | Service Level | Cost red. MRP stand. |
| 1 | 0.02 | 0.0 | 2 | FOP 1 | 6,063 | 71% | 0% | 0.2 | 3 | FOQ 200 | 7,355 | 98% | -26% ** |
| 1 | 0.04 | 0.0 | 3 | FOQ 200 | 7,936 | 98% | 0% | 0.4 | 3 | FOQ 200 | 8,366 | 98% | -34% ** |
| 1 | 0.06 | 0.2 | 3 | FOP 1 | 8,762 | 97% | -14% ** | 0.4 | 3 | FOQ 200 | 8,922 | 94% | -34% ** |
| 1 | 0.08 | 0.2 | 4 | FOQ 200 | 11,165 | 97% | -19% ** | 0.6 | 4 | FOQ 200 | 11,284 | 98% | -20% ** |
| 1 | 0.10 | 0.4 | 4 | FOQ 200 | 13,164 | 93% | -14% ** | 0.6 | 4 | FOQ 200 | 12,751 | 92% | -26% ** |
| 1 | 0.12 | 0.4 | 4 | FOQ 200 | 14,767 | 86% | -13% ** | 0.8 | 4 | FOQ 200 | 16,088 | 86% | -21% * |

| | | High Utilization MRP extended | | | | | | | | | | | |
|---|---|---|---|---|---|---|---|---|---|---|---|---|---|
| | | Temporary Overbooking | | | | | | Temporary Underbooking | | | | | |
| β | α | SST | PLT | Lotsize | Overall Costs [CU] | Service Level | Cost red. MRP stand. | SST | PLT | Lotsize | Overall Costs [CU] | Service Level | Cost red. MRP stand. |
| 1 | 0.02 | 0.0 | 3 | FOP 1 | 6,804 | 100% | 0% | 0.4 | 3 | FOQ 200 | 7,930 | 100% | -20% ** |
| 1 | 0.04 | 0.2 | 3 | FOP 1 | 8,646 | 97% | -4% | 0.4 | 3 | FOQ 200 | 8,710 | 92% | -43% ** |
| 1 | 0.06 | 0.2 | 4 | FOP 2 | 10,693 | 93% | -19% ** | 0.6 | 4 | FOQ 200 | 11,608 | 95% | -29% ** |
| 1 | 0.08 | 1.5 | 1 | FOQ 200 | 14,036 | 95% | -7% * | 1.5 | 1 | FOQ 200 | 14,090 | 92% | -20% ** |
| 1 | 0.10 | 1.5 | 1 | FOQ 800 | 16,100 | 95% | -5% | 1.5 | 1 | FOQ 800 | 16,443 | 94% | -14% * |
| 1 | 0.12 | 1.5 | 2 | FOQ 800 | 17,585 | 96% | -5% | 2.0 | 1 | FOQ 800 | 18,740 | 95% | -19% ** |

## 7. Conclusion

In this paper, the influence of uncertain forecast demand updates on the optimal planning parameters for MRP is investigated and an *MRP safety stock exploitation heuristic* is developed. In addition to unsystematic uncertainties in demand forecasts, the effect of biased forecasts is also evaluated. Simulation is applied to model the planning and production system structure and evaluate overall costs, i.e., the sum of inventory and backorder costs.

The study identifies limitations within the standard MRP approach, particularly how the benefits of improved demand information are offset by increased system nervousness due to forecast updates. In scenarios of high forecast uncertainty, standard MRP parameters tend to replicate a make-to-stock policy, failing to effectively leverage uncertain demand forecast information. Conversely, our developed *MRP safety stock exploitation heuristic* demonstrates robust performance across test instances, significantly reducing costs by effectively utilizing uncertain demand information. This is particularly evident in overbooking scenarios, where our heuristic significantly lowers costs, supporting the observations of Enns (2002) as well as Sridharan and Lawrence LaForge (1989) that overestimating forecasts or moderate hedging of the MPS is beneficial. Moreover, our developed heuristic not only diminishes costs but also stabilizes production system fluctuations, thereby reducing MRP nervousness compared to the method proposed by Sridharan and Lawrence LaForge (1989). It also highlights the necessity of a capacity buffer to maximize the heuristic's effectiveness, aligning with Wijngaard (2004). From a managerial perspective, a key insight is that underestimating demand forecasts poses a greater risk to manufacturing companies than overestimation. Furthermore, another interesting managerial insight is that, especially in overbooking scenarios with more accurate forecasts (i.e., lower uncertainty), FOP lot-sizing should be preferred, while in unbiased forecast settings, FOQ lot-sizing almost always performs better.

The study shows for RQ1 that, for both FOP and FOQ lot-sizing policies, overall costs at optimal MRP planning parameters more than double when comparing the lowest to the highest uncertainty levels, regardless of whether demand updates are biased or unbiased, when applying the *MRP standard*. The study demonstrates that, in response to RQ2, the *MRP safety stock exploitation heuristic* reduces total costs, compared to the MRP standard, by an average of 21% for unbiased demand updates. In response to RQ3, the heuristic reduces costs by 9% for permanent overbooking, 27% for permanent underbooking, 8% for temporary overbooking, and 26% for temporary underbooking. These cost reductions were computed as averages across all utilization and uncertainty levels tested. The cost reductions are statistically significant in most scenarios, even at $\alpha=0.01$, confirming the heuristic's effectiveness in improving MRP performance. However, this significance diminishes in overbooking scenarios with high utilization and high forecast uncertainty.

Based on the results discussion above, further research could address into more complex interrelations between demand forecasts, i.e., correlations between products and autocorrelation between demand periods, automatic corrections of demand forecast bias, and more intricate production systems including scrap. Additionally, exploring advanced optimization algorithms (e.g., hybrid heuristics and metaheuristics, adaptive algorithms, self-adaptive algorithms, island algorithms, polyploid algorithms) could offer significant insights for enhancing MRP systems to better capitalize on uncertain demand forecasts. This necessitates a broader discussion on the importance of these sophisticated optimization techniques in tackling challenging decision problems.

## ACKNOWLEDGEMENTS

This research was funded in whole, or in part, by the Austrian Science Fund (FWF) [10.55776/P32954]. For the purpose of open access, the author has applied a CC BY public copyright license to any Author Accepted Manuscript version arising from this submission.

**DATA AVAILABILITY STATEMENT**

The data that support the findings of this study are available in Zenodo at https://doi.org/10.5281/zenodo.7963040 upon reasonable request to the author, Altendorfer K.